\documentclass[aps,showpacs,twocolumn]{revtex4}
\usepackage{graphics,epsfig,epsf,psfrag}

\usepackage{amsmath,amssymb,mathrsfs,latexsym}
\usepackage{slashed}
\usepackage{bbm}
\usepackage[usenames]{color}

\newcommand{\Lag}{\mathcal{L}}

\newcommand{\Tr}{\textrm{Tr}}

\newcommand{\JJ}{\mathcal{J}}

\begin{document}

\title{Search for $J^{PC}=1^{-+}$ exotic state in $e^+e^-$ annihilation}

\author{ Qian Wang\footnote{{\it E-mail
address:} q.wang@fz-juelich.de}}

\affiliation{
       Institut f\"{u}r Kernphysik, Institute for Advanced Simulation and J\"ulich Center for Hadron
          Physics, Forschungszentrum J\"{u}lich, D-52425 J\"{u}lich, Germany}

\begin{abstract}
The charmed meson pair $(\frac{3}{2})^+$ and anti-$(\frac{1}{2})^-$, i.e. $D_1(2420)\bar D+c.c.$, $D_1(2420)\bar D^*+c.c.$, $D_2(2460)\bar D^*+c.c.$, can couple to states with vector
quantum number $J^{PC}=1^{--}$ and exotic quantum number $J^{PC}=1^{-+}$ in a relative $S$ wave. Near threshold, the charmed meson pair may form hadronic molecules due to the strong $S$-wave
coupling, and the mysterious vector state $Y(4260)$ could be such a state of the $D_1(2420)\bar D+c.c.$ molecule. This implies the possible existence of its exotic partner made of the same charmed mesons but with $J^{PC}=1^{-+}$. We evaluate the production rate of such exotic hadronic molecules and propose a direct experimental search for them in $e^+e^-$
annihilation. The confirmation of such exotic states in experiment will shed light on the spectrum in the heavy quark sector. 

\end{abstract}

\date{\today}

%\pacs{14.40.Rt, 13.75.Lb, 13.20.Gd}
\pacs{}

\maketitle

Quantum Chromodynamics (QCD) allows for a much richer and more complicated multi-quark systems beyond the simple quark model where mesons and baryons are color singlets made of $q\bar q$ and
$qqq$, respectively. This reflects the complexity related to structure formation in the strong interactions, which is still an unsolved question in fundamental science. Addtionally,
a discovery of hadron states beyond the simple quark model will greatly extend our knowledge of the strong force that  binds the matter in our universe.

During the past few years there has been important experimental progress in the search for exotic hadrons. The Belle Collaboration reported observations of charged bottomonium states,
$Z_b^{(\prime)\pm}$\cite{Belle:2011aa,Adachi:2012cx}, and last year BESIII reported their analogues in the charm sector, i.e. the charged charmonium states
$Z_c^{(\prime)\pm}$\cite{Ablikim:2013mio,Liu:2013dau,Ablikim:2013emm,Ablikim:2013wzq}. The BESIII observation of the $Z_c(3900)$ was soon confirmed by the Belle
Collaboration~\cite{Liu:2013dau} and a re-analysis based on the CLEO-c data~\cite{Xiao:2013iha}. These observations immediately initiated numerous studies of their intrinsic structure, production and decay mechanisms. In particular, the confirmation of the $Z_c(3900)$ helps us recognize that the $S$-wave thresholds of heavy quark meson pairs play an essential role in the formation of these charged mesons with hidden heavy quarkonium.  Particularly, the production mechanism of the $Z_c(3900)$ may shed light on the role played by the $S$-wave open charm thresholds in the vector sector and the nature of the initial vector meson $Y(4260)$. 

The lowest open charm thresholds in the vector sector are created by the low-mass $D$ and/or $D^*$ pairs. They couple to the vector quantum number $J^{PC}=1^{--}$ via a $P$
wave. Since it is much harder to form a P-wave molecule than to form a S-wave molecule, there may be some S-wave vector molecules which are much related to the corresponding thresholds. Meanwhile the first S-wave open-charm threshold is  the $D_1(2420)\bar D +c.c.$ which is located at 4.29~GeV.  Recently, due to the observation of $Z_c(3900)$, the role of the $S$-wave open charm thresholds in the vector meson production via the $e^+e^-$ annihilations has been carefully studied~\cite{Wang:2013cya,Wang:2013kra,Wang:2013hga}.  As pointed
out in our recent works, the mysterious $Y(4260)$ can be a manifestation of the $S$-wave $D_1(2420)\bar D +c.c.$ threshold since its wave function might be dominated by a molecular $D_1(2420)\bar D +c.c.$ component~\cite{Wang:2013cya,Wang:2013kra,Wang:2013hga}.  Near threshold, the  production of the $S$-wave $(\frac{3}{2})^++$ anti-$(\frac{1}{2})^-$ meson pairs is more important than that of $(\frac{1}{2})^-+$ anti-$(\frac{1}{2})^-$ meson pairs.  In the heavy quark limit, the heavy mesons can be classified by their light degrees of freedom, i.e. $s_l=l\pm \frac 12$. So the charmed mesons $(D, D^*)$ and $(D_1, D_2)$ belong to the $(\frac 12)^-$ multiplet and the $(\frac 32)^+$ multiplet respectively.  Although the production of   $(\frac{3}{2})^++$ anti-$(\frac{1}{2})^-$ heavy meson pairs is suppressed in the heavy quark
limit~\cite{Li:2013yka}, the heavy quark spin symmetry breaking effects in the charm sector can be significant as discussed in Ref.~\cite{Wang:2013kra}. The molecular prescription of the $Y(4260)$ is consistent with its properties observed in experiment so far and provides a natural explanation for the production of the $Z_c(3900)$.  Furthermore, it opens a door towards a dynamical study of hadronic molecules with exotic quantum numbers, i.e. $J^{PC}=1^{-+}$, based on the
heavy quark spin symmetry.

In this letter, we focus on the production of the $(\frac{3}{2})^++$ anti-$(\frac{1}{2})^-$ molecules with quantum number $1^{-+}$. We analyse their correlation with the $1^{--}$ partners in
the non-relativistic effective field theory (NREFT) and discuss their electromagnetic transitions from the $1^{--}$ vector charmonia. This suggests a direct search for their signals at
$e^+e^-$ colliders.

As studied in Ref.~\cite{Wang:2013cya}, the recently observed $Z_c(3900)$~\cite{Ablikim:2013mio,Liu:2013dau} can be explained naturally in the $D_1D$ (here, $D_1D$ means $D_1\bar D+c.c.$. The similar conventions $D_1D^*$ and $D_2D^*$ are applied for $D_1\bar D^*+c.c.$ and $D_2\bar D^*+c.c.$, respectively) molecular picture for the $Y(4260)$, since a large number of $D\bar D^*+c.c.$ meson pairs can be produced via $D_1\to D^*\pi$.  In the further study of the $Y(4260)$ line shape in Ref.~\cite{Cleven:2013mka},
we extend this scenario to include the $D_1D$, $D_1D^*$ and $D_2D^*$ thresholds to explain its line shapes in $J/\psi\pi\pi$ and $h_c\pi\pi$ channels.  That means if $Y(4260)$ is the     $D_1D$ vector molecule, it is also possible that the other $D_1D^*$ and $D_2D^*$ vector molecules exist. In this letter, these three vector molecules are denoted by $Y$, $Y^\prime$ and $Y^{\prime\prime}$ respectively.  The Lagrangians between these three vector molecules and their components reads
\begin{eqnarray}
\hspace{-0.5cm}
  \Lag_{Y}&=&\frac{y}{\sqrt{2}} Y^{i} \left( D_{1a}^{i\dag} \bar D_a^\dag-D_a^\dag
\bar D_{1a}^{i\dag} \right) \nonumber\\
&+ &i\frac{y^\prime}{\sqrt 2}\epsilon^{i j k}Y^{\prime i}\left( \bar D_1^{k\dag }D^{*j\dag} - D_1^{k\dag }\bar D^{*j\dag}\right) \nonumber\\
&+ &\frac{y^{\prime\prime}}{\sqrt 2} Y^{\prime\prime i}\left( \bar D_2 ^{ij\dag}D^{*j\dag} - D_2 ^{ij\dag}\bar D^{*j\dag}\right)+ \text{H.c.}
\label{eq:Lag1}
\end{eqnarray}
Here, we assume that their charged partners also exist, i.e. the $1^{-+}$ molecules denoted as $X$, $X^\prime$ and $X^{\prime\prime}$ respectively, for which the relevant Lagrangian is
\begin{eqnarray}
\hspace{-0.5cm}
  \Lag_{X}&=&\frac{x}{\sqrt{2}} X^{i} \left( D_{1a}^{i\dag} \bar D_a^\dag+D_a^\dag
\bar D_{1a}^{i\dag} \right) \nonumber\\
&+ &i\frac{x^\prime}{\sqrt 2}\epsilon^{i j k}X^{\prime i}\left( \bar D_1^{k\dag }D^{*j\dag}
+ D_1^{k\dag }\bar D^{*j\dag}\right) \nonumber\\
&+ &\frac{x^{\prime\prime}}{\sqrt 2} X^{\prime\prime i}\left( \bar D_2 ^{ij\dag}D^{*j\dag} + D_2 ^{ij\dag}\bar D^{*j\dag}\right)+ \text{H.c.}
\label{eq:Lag2}
\end{eqnarray}
The above couplings and their relative phases should, in principle, be determined by  detailed  dynamics.
At this moment, since we lack such constraints on the couplings we then assume that these three states, $X$, $X^\prime$ and $X^{\prime\prime}$, are dominated by the $D_1D$, $D_1D^*$ and $D_2D^*$ components, respectively. Nevertheless, we expect that their masses are not far away from their respective thresholds which would allow the implementation of the Weinberg
criterion~\cite{Guo:2013zbw,Weinberg:1965zz}. As an example for the $X$ state as the $D_1D$ molecule, we have
\begin{equation}
  \label{eq:effectivecoupling}
  x^2 = \lambda^2 \frac{16\pi}{\mu}
\sqrt{\frac{2\epsilon}{\mu}}
\left[ 1 + \mathcal{O} \left( \sqrt{2\mu \epsilon}\,r \right) \right],
\end{equation}
with $\mu$ the reduced mass, $r$ the range of the force, $\epsilon$ the binding energy, and $\lambda$ the probability to find their components in the physical wave functions. For a pure
bound states, $\lambda=1$~\cite{Guo:2013zbw}.  Eq.(\ref{eq:effectivecoupling}) can also be applied to the other five coupling constants, i.e. $x^\prime$, $x^{\prime\prime}$, $y$, $y^\prime$ and $y^{\prime\prime}$. 

In the molecular scenario, the radiative transitions between the $1^{--}$ and $1^{-+}$ states are through their constituents, namely the charmed mesons. The corresponding effective
Lagrangians for the radiative transitions are
\begin{eqnarray}\nonumber
\Lag_{HH\gamma} = \frac{e\,\beta}{2} \Tr\left[ H_a^\dag H_b\,
\vec{\sigma}\cdot \vec{B} \, Q_{ab} \right] + \frac{e\, Q'}{2m_Q} \Tr \left[
H_a^\dag \,
\vec{\sigma}\cdot \vec{B} \, H_a\right]
\label{eq:LHHga}
\end{eqnarray}
and
\begin{eqnarray}\nonumber
 \Lag_{TT\gamma}=\frac{e \beta^\prime}{2}\Tr \left[ T_a^{ k\dag} T_b^{k} \vec\sigma \cdot \vec B Q_{ab}\right]+\frac{e Q^\prime}{2 m_Q}\Tr \left[ T_a^{ k\dag} \vec\sigma \cdot \vec B T_a^{k} \right]
 \label{eq:LTTga}
 \end{eqnarray}
with $B^k=\varepsilon^{ijk}\partial^iA^j$ the magnetic field, $Q_{ab}=\mathrm{diag}(2/3,-1/3,-1/3)$ the matrix of the light quark charge and $Q'$ the heavy quark charge (in units of the
proton charge $e$).
The parameters $\beta^{-1}=276~\mathrm{MeV}$ and $m_c=1.5~\mathrm{GeV}$ are taken from Ref.~\cite{Hu:2005gf}. The details of $H$ and $T$ can be found in Ref.~\cite{Guo:2013zbw}.

Since the spin structure of the $(\frac 32)^+$ multiplet is more complicated than that of the $(\frac 12)^-$ multiplet, there are more degrees of freedom for the $(\frac 32)^++(\frac 12)^-$ system than that for the $(\frac 12)^-+(\frac 12)^-$ system~\cite{Guo:2013sya,Valderrama:2012jv,Nieves:2012tt}.  That means, in the $(\frac 32)^++(\frac 12)^-$  system, we could not expect all the potentials are universal  for different quantum numbers as well as for different constituents. However we could assume their potentials for different quantum numbers and different constituents do not vary too much.  That means we could expect the binding energies of these six molecules would be of the same order if they exist as bound states. To simplify this issue, we assume they are the same.
 Using the results of the fit in Ref.~\cite{Cleven:2013mka}, we roughly estimate the binding energy is $\epsilon\sim 70~\mathrm{MeV}$. 
As a result, the masses of the corresponding
$D_1D^*$ and $D_2D^*$ molecules are
\begin{eqnarray}\label{mass-2Y}
M_{Y^\prime(X^\prime)}\sim 4361~\mathrm{MeV},\quad M_{Y^{\prime\prime}(X^{\prime\prime})}\sim 4403~\mathrm{MeV}.
\end{eqnarray}
The candidates for these two vector states could be $Y(4360)$ and $\psi(4415)$. Since the width of $D_2$ is as twice large as that of $D_1$, the signal of the $D_2D^*$ threshold effect near the  $Y^{\prime\prime}$ is not as significant
as the other two thresholds~\cite{Wang:2013hga}. To confirm or exclude the $D_2D^*$ molecular interpretation of $\psi(4415)$, the angular distribution analysis of $DD^*\pi$ and $D^*D^*\pi$ at $4.415~\mathrm{GeV}$ is needed.

In the vector sector the mass region of 4$\sim$4.7~GeV is far from well understood. Although the $\psi(4040)$, $\psi(4160)$ and $\psi(4415)$ are assigned as conventional $3S$, $2D$ and $4S$
vector charmonia, the $Y(4260)$ and newly observed $Y(4360)$ and $Y(4660)$ cannot be well organized. Various explanations were proposed in the literature to explain the nature of these
states. In Ref.~\cite{Segovia:2013wma}, the $Y(4360)$ and $\psi(4415)$ are classified as conventional $4S$ and $3D$ vector charmonia in the constituent quark model, while in
Ref.~\cite{Ding:2007rg} they are interpreted as the canonical $3^3D_1$ and $5^3S_1$ states.  Recently, Li and Voloshin argue that $Y(4260)$ and $Y(4360)$ are the mixing of spin-singlet and
spin-triplet hadro-charmonia~\cite{Li:2013ssa}. In Ref.~\cite{Guo:2008zg} the $Y(4660)$ is regarded as the $\psi^\prime f_0(980)$ molecule. In contrast, we propose that the $Y(4260)$,
$Y(4360)$ and $\psi(4415)$ are most likely to be the $D_1D$, $D_1D^*$ and $D_2D^*$ vector molecules~\cite{Cleven:2013mka}, which is consistent with the estimate of Eq.~(\ref{mass-2Y}). However, their analogues in the bottom sector cannot be studied in $e^+e^-$ colliders, since their production is highly suppressed in the heavy quark limit~\cite{Li:2013yka}.

\begin{figure}[htbp]
\centering
\includegraphics[width=0.6\linewidth]{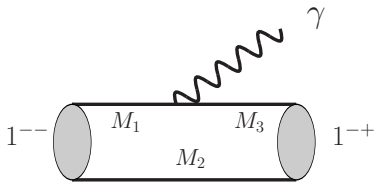}
\caption{The schematic diagram of the radiative transition between $(\frac{3}{2})^++(\frac{1}{2})^-$ molecules with quantum number $1^{--}$ and $1^{-+}$. $M_1$, $M_2$ and $M_3$ are the constituents of the corresponding molecules.}
\label{fig:Feynmann}
\end{figure}

\begin{table*}[htbp]
\begin{center}
\caption{The amplitudes of the radiative transitions between different $1^{--}$ and $1^{-+}$ molecular states.  The symbol  ``--"  means that the transition is forbidden at leading order.}
\vspace{0.5cm}
\begin{tabular}{|c|c|c|c|}
\hline\hline
 $1^{--}\to 1^{-+}\gamma$ & $X$ & $X^{\prime}$ & $X^{\prime\prime}$ \\
 \hline
 $Y$ &$[D_1DD_1]$ & $[DD_1D^*]$ &--  \\\hline
$Y^\prime$ & $[D^*D_1D]$ & $[D_1D^*D_1], [D^*D_1D^*]$ & $[D_1D^*D_2]$  \\\hline
 $Y^{\prime\prime}$ & -- & $[D_2D^*D_1]$ & $[D_2D^*D_2], [D^*D_2D^*]$ \\\hline\hline
 \end{tabular}
 \label{tab:amplitudetransition}
 \end{center}
 \end{table*}
The radiative transitions between heavy quarkonia  provide an insight into the structure of the heavy quark bound states. It also offers us a method to explore the nature of the hadronic
molecules. Radiative transitions between different molecules are through their constituents, such as $D^*\to D\gamma$ through the M1 transition and $D_1\to D_1\gamma$ from the minimal substitution in kinematic term and the M1 transition in Eq.(\ref{eq:LTTga}).  The schematic diagram illustrating the radiative transition between the $1^{--}$ and $1^{-+}$ molecular states is
shown in Fig.~\ref{fig:Feynmann}.  $[M_1M_2M_3]$ is used to denote different diagrams.  The possible radiative transitions between the Y-type  and X-type molecules are listed in Table~\ref{tab:amplitudetransition}. The mesons in the squared brackets are the intermediate
mesons in the scalar three-point loop function as defined by Eq.~(A.1) in Ref.~\cite{Cleven:2013sq}. There is only one Lorentz structure $\vec\epsilon_X\cdot\hat k
\vec\epsilon_\gamma\cdot\vec\epsilon_Y-\vec\epsilon_Y\cdot\hat k \vec\epsilon_\gamma\cdot\vec\epsilon_X$ with $\vec\epsilon_X$, $\vec\epsilon_Y$ and $\vec\epsilon_\gamma$ the polarization
vectors of the corresponding $X$, $Y$ and photon, respectively, and $\hat{k}$ the unit vector along the three momentum of photon.

The radiative transitions among the Y-type and X-type molecules at leading order are listed in  Table~\ref{tab:amplitudetransition}. The radiative transition between $Y$ and $X^{\prime\prime}$, $Y^{\prime\prime}$ and $X$ can only happen when the next leading order diagrams are considered.  As discussed in the above paragraph, all the
amplitudes have the same Lorentz structure with additional coupling constants and a three point scalar loop function.
Due to the transverse property of the vector meson produced in $e^+e^-$ collider, the angle between the photon momentum and the polarisation vector $\vec{\epsilon}_Y$ can be related to that relative to the beam axis. So the angular distribution of the process $Y^{(\prime,\prime\prime)}\to X^{(\prime,\prime\prime)}\gamma$ is 
\begin{eqnarray}
\frac{d\Gamma}{d \cos\theta}\propto 1+\frac{3}{2}\sin^2\theta,
\label{eq:distribution}
\end{eqnarray}
with $\theta$ the angule of photon momentum relative to the beam axis. The equation
\begin{eqnarray}
  \overline {\sum_{\lambda=1,2}} \left| \hat k \cdot
\vec\epsilon_{Y^{(\prime,\prime\prime)}}^{\,(\lambda)}
\right|^2 = \frac12 \sin^2\theta,
\end{eqnarray}
is used to deduce Eq.~(\ref{eq:distribution}), where the ``---" denotes averaging the polarisations of the $Y^{(\prime,\prime\prime)}$. The angular distribution in Eq.~(\ref{eq:distribution}) is very different from that of $Y\to X(3872)\gamma$~\cite{Guo:2013zbw}, and can be
easily detected by experiments with sufficiently large statistics.
\begin{figure}[htbp]
\centering
\includegraphics[width=0.8\linewidth]{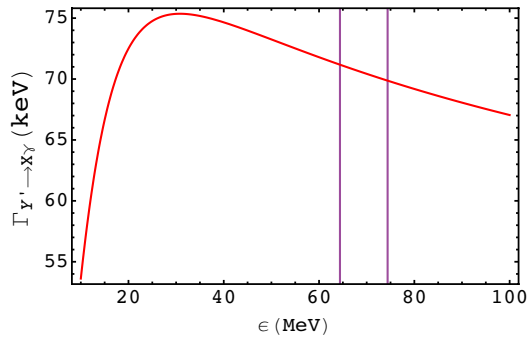}
\caption{The partial width of $Y^\prime\to X+\gamma$ as a function of the binding energy is shown. The two vertical lines are the lower limit and upper limit
of the binding energy from the fit in Ref~\cite{Cleven:2013mka}. }
 \label{fig:partialwidth}
\end{figure}

Since the phase space of $Y^\prime\to X\gamma$ is as twice large as that of $Y^{\prime\prime}\to X^\prime\gamma$, it has the largest probability to be detected by experiment. In the following, we study the radiative decay of $Y^\prime$ to $X$. 
With the width of $D_1$ being taken into account,  the partial width of $Y^\prime\to X\gamma$ with respect to the binding energy is shown Fig.~\ref{fig:partialwidth}.  Considering the
binding energy of roughly $70~\mathrm{MeV}$, we obtain the partial width $\Gamma(Y^\prime\to X\gamma)\sim 70 ~\mathrm{keV}$ and the branching ratio $BR(Y^\prime\to X\gamma)\sim 10^{-3}$ with $\Gamma(Y(4360))=74~\mathrm{MeV}$~\cite{Beringer:1900zz}.

Using Eq.~(\ref{eq:effectivecoupling}) with the binding energy $70~\mathrm{MeV}$, we obtain the coupling constants in Eq.(\ref{eq:Lag1}) and Eq.(\ref{eq:Lag2})
\begin{eqnarray}
y(x)=2.35~\mathrm{GeV}^{\frac 12},\quad y^\prime(x^\prime)=2.28~\mathrm{GeV}^{\frac 12}.
\end{eqnarray}
Following the same procedure as that in Refs.~\cite{Cleven:2013mka,Wu:2013onz}, the cross section of the full process, i.e. $e^+e^-\to Y(Y^\prime)\to X\gamma$, can be extracted as
\begin{eqnarray}\nonumber
\sigma(s)=(4\pi\alpha)^2\left (g_{\gamma^*Y(Y^\prime)}\frac{M_Y^2}{s}\right)^2(M_Y\Gamma_{Y(Y^\prime)\to X\gamma})|G_Y(s)|^2 \ ,
\label{eq:crosssection}
\end{eqnarray}
where $g_{\gamma^*Y(Y^\prime)}$ is the dimensionless coupling constant between the virtual photon and vector state $Y (Y^\prime)$, and $G_Y(s)$ is the corresponding
propagator~\cite{Cleven:2011gp}
\begin{eqnarray}
G_Y^{-1}=s-M_Y^2+\hat\Pi\left(s\right)+iM_Y\Gamma_Y
\label{eq:Prop}
\end{eqnarray}
with
\begin{eqnarray}\nonumber
\hat\Pi\left(s\right)=\Pi\left(s\right)-\mathrm{Re}\left[\Pi\left(M_Y^2\right)+\left(s-M_Y^2\right)\partial_{s}\Pi(s)|_{s=M_Y^2}\right] \ .
\end{eqnarray}
Here we assume the constant widths of $Y (Y(4260))$ and $Y^\prime (Y(4360))$ in Eq.~(\ref{eq:Prop}) are the same, i.e. $\Gamma=40~\mathrm{MeV}$~\cite{Cleven:2013mka}.  Since both $Y$ and $Y^\prime$ can
decay to $X\gamma$, the relative strength will depend on the coupling constant $\beta^\prime$.   In the following, we assume that the coupling $\beta^\prime=\beta (1\pm
0.5)$. 
\begin{figure}[htbp]
\centering
\includegraphics[width=0.8\linewidth]{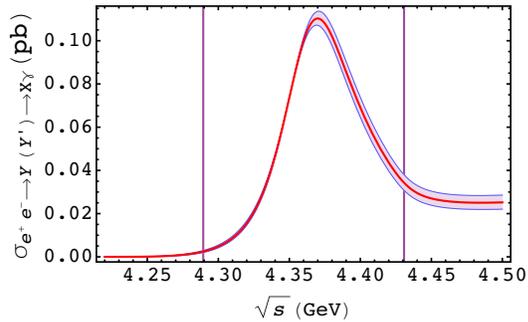}
\caption{The cross section of $e^+e^-\to Y(Y^\prime)\to X \gamma$ in terms of the centre energy is illustrated. The vertical lines are the corresponding $D_1D$ and $D_1D^*$ thresholds.  The boundary corresponds to the $50\%$ uncertainty of the coupling constant $\beta^\prime$. }
\label{fig:CrossSection}
\end{figure}
As a result, the cross section of $e^+e^-\to Y(Y^\prime)\to X\gamma$ in the energy region [4.20, 4.50]~GeV is shown in Fig.~\ref{fig:CrossSection} and turns out to be nontrivial. As discussed before, the radiative transition between  $Y^\prime$ and $X$ is more important than that between $Y$ and $X$ which makes the boundary from the uncertainty of $\beta^\prime$ negligible. Although the cross
section of 0.1 pb around 4.36~GeV is smaller than those of $J/\psi\pi\pi$ and $h_c\pi\pi$, which are roughly $70\sim 80$ pb, it is still larger than that of the isospin violating process
$J/\psi\eta\pi^0$~\cite{Wu:2013onz}. It can be possibly measured at BESIII and help us understand the nature of these exotic states.

In the molecular scenario, i.e. $(\bar Q\Gamma^A q) (\bar q \Gamma^B Q)$ with $Q$ the heavy quark and $q$ the light quark, we can study their hidden charmonium decay patterns by using Fierz transformation~\cite{Bondar:2011ev}  as well as
the 9-j symbol method~\cite{Li:2013yka}.  Here we employ the Fierz transformation 
\begin{eqnarray}
(\bar u_1\Gamma^A u_2)(\bar u_3 \Gamma^B u_4)=\sum_{CD}C^{AB}_{CD}(\bar u_1\Gamma^C u_4) (\bar u_3 \Gamma^D u_2)
\label{eq:Fierz}
\end{eqnarray}
with $\Gamma^i=(\bf1, \gamma_\mu, \frac i 2 [\gamma_\mu,\gamma_\nu], \gamma_\mu\gamma_5, i \gamma_5)$ and $C^{AB}_{CD}=\frac {1} {16} \mathrm{Tr}[\Gamma^C\Gamma^A\Gamma^D\Gamma^B]$ to study their hidden charm decay modes. The currents $\bar q \gamma_\mu\gamma_5 c$, $\bar c \gamma_\mu\gamma_5 q$, $ i \bar q \gamma_5 c$, $i \bar c\gamma_5 q$ describe $D_1$, $\bar D_1$, $D$,  $\bar D$ respectively.  When one analyses their hidden charm decay patterns, the two currents of its constitutes must be located at the same position.  That is the reason why we can use Fierz transformation to study their hidden charm decays qualitatively. 
\begin{table}
\begin{center}
\caption{The mainly hidden charm decay modes of the $1^{--}$ and $1^{-+}$ molecular states are listed. $\JJ_{q,j}$ means the other light degrees of freedom with their corresponding quantum number $j$ in the sub-index.}
\vspace{0.5cm}
\begin{tabular}{|c|c|c|c|}
\hline\hline
 constituents & $D_1D$ & $D_1D^*$ & $D_2D^*$ \\
 \hline
 $1^{--}$ &$J/\psi+\JJ_{q,S}$ & $J/\psi+\JJ_{q,V}$ &$J/\psi+\JJ_{q,S}$  \\\hline
$1^{-+}$ & $\eta_c+\JJ_{q,A}$ & $\eta_c+\JJ_{q,T}$ & $\eta_c+\JJ_{q,A}$  \\\hline\hline
 \end{tabular}
 \label{tab:amplitude}
 \end{center}
 \end{table}
As shown in Table~\ref{tab:amplitude}, we find that the dominant hidden charm decay modes of all the Y-type molecules are $J/\psi$ plus some light mesons.
That would explain why the $Y(4260)$ is only well established in the $J/\psi\pi\pi$ channel until now,
while in the $h_c\pi\pi$ channel, there are only indications of the $Y(4260)$. Meanwhile, the X-type molecules mainly decay to $\eta_c$ plus some light mesons, which should be useful for future experimental detections.

In this letter,  we study the radiative transitions among the Y-type molecules and X-type molecules in the molecular picture.  Due to the large phase space, the partial width of $Y^\prime\to X\gamma$ would be one of the largest ones. We estimate its partial width is $\Gamma(Y^\prime\to X\gamma)\sim 70~\mathrm{keV}$ giving the
branching ratio $BR(Y^\prime\to X\gamma) \sim 10^{-3}$ which could be tested by BESIII further. At the same time, their angular distribution, i.e. $1+\frac{3}{2}\sin^2\theta$, is another quantity which could help us clarify their nature.
 As a simple extension, we estimate the cross section of $e^+e^-\to Y(Y^\prime)\to X\gamma$ is about $0.1$~pb at around 4.36~GeV. Although the line shape has some uncertainty from the coupling
 $\beta^\prime$,
the trend could give some guidance for the experimentalists. For their hidden charm decay modes, the Y-type molecules mainly decay to $J/\psi$ plus some light mesons and the X-type molecules mainly decay to $\eta_c$ plus some light mesons. To pin down their natures, high statistic data from BESIII are needed in future.  
\medskip

%\begin{acknowledgments}
 Special thanks to Christoph Hanhart, Ulf-G. Mei{\ss}ner and  Qiang Zhao for useful comments and suggestions on improving this work, and to Feng-Kun Guo for useful discussions and for having benefited from the ``AmpCalc.m" package written by him. Useful discussions with  Wei Wang are also acknowledged.
This work is supported by DFG and NSFC through funds provided to the Sino-German CRC 110 ``Symmetries and the Emergence of Structure in QCD''.

%\end{acknowledgments}

\end{document}